\documentclass[12pt]{iopart}

\usepackage{iopams}
\usepackage{graphicx}
\begin{document}

\title[Cluster mean-field theory study of $J_{1}-J_{2}$ Heisenberg model on a square lattice]{Cluster mean-field theory study of $J_{1}-J_{2}$ Heisenberg model on a square lattice}

\author{Yong-Zhi Ren$^1$, Ning-Hua Tong$^1$ and Xin-Chen Xie$^2$}

\address{$1$ Department of Physics, Renmin University of China,
100872 Beijing, China}
\address{$2$ International Center for Quantum Materials and School
of Physics, Peking University, Beijing 100871, China}
\ead{renyongzhi@ruc.edu.cn}

\begin{abstract}
We study the spin-$1/2$ $J_{1}$-$J_{2}$ Heisenberg model on a square
lattice using the cluster mean-field theory. We find a rapid
convergence of phase boundaries with increasing cluster size. By
extrapolating the cluster size $L$ to infinity, we obtain accurate
phase boundaries $J_{2}^{c1} \approx 0.42$ (between the
N$\acute{e}$el antiferromagnetic phase and nonmagnetic phase), and
$J_{2}^{c2} \approx 0.59$ (between nonmagnetic phase and the
collinear antiferromagnetic phase). The transitions are identified
unambiguously as second order at $J_{2}^{c1}$ and first order at
$J_{2}^{c2}$. At finite temperature, we present a complete phase
diagram with stable, meta-stable and unstable states near
$J_{2}^{c2}$, being relevant to that of the anisotropic $J_1-J_2$
model. The uniform as well as staggered magnetic susceptibilities
are also discussed.

\noindent{\it Keywords\ $J_{1}-J_{2}$ Heisenberg model, quantum
phase transition, cluster mean-field theory }
\end{abstract}

\maketitle

\section{Introduction}
It was suggested by P. W. Anderson\cite{Anderson} that low spin, low
spatial dimension, and high frustration are the three main factors
which favor the melting of magnetic long range order (LRO) and lead
to exotic spin liquid ground state. Such a state was closely related
to the appearance of superconductivity in the high-temperature
superconductivity in Cu-based oxides upon doping\cite{Review}. The
spin-$1/2$ $J_{1}$-$J_{2}$ Heisenberg model in two dimensional
square lattice is such a model that bears all the three factors,
hence its ground state is a promising candidate for the exotic spin
liquid state\cite{Chandra1}. Besides the interest for spin liquid,
this model in the large $J_2/J_1$ regime is relevant to materials
such as $Li_2VOSiO_4$\cite{Melzi}, and the $S> 1/2$ version is
relevant to the parent material of iron-based high temperature
superconductors\cite{Yildirim}.

The Hamiltonian of antiferromagnetic (AFM) $J_{1}-J_{2}$ model reads
\begin{equation}
    \hat{H}=J_{1}\sum_{\langle i,j \rangle}\mathbf{S}_{i}\cdot\mathbf{S}_{j}
    + J_{2} \sum_{\langle\langle
    i,j\rangle\rangle} \mathbf{S}_{i} \cdot \mathbf{S}_{j},
\end{equation}
where $\mathbf{S_{i}}$ is the spin $\frac{1}{2}$ operator on site
$i$, $J_{1}$ and $J_{2}$ are the nearest neighbor and the
next-nearest neighbor coupling coefficients, respectively. In the
following, we set $J_{1}=1$ as the unit of energy. For the
next-nearest neighbor coupling $J_{2}$, we confine ourself to the
AFM case $J_{2}>0$.

This model received numerous studies in the past two decades, using
various methods including exact diagonalization
(ED)\cite{Dagotto,Schulz, Richter,Capriotti1,Mambrini}, series
expansion\cite{Gelfand1,Gelfand2,Singh1,Singh2,Sirker}, coupled
cluster\cite{Schmalfuss, Darradi}, spin wave
approximation\cite{Chandra1,Dotsenko}, Green's function
method\cite{Siurakshina}, density-matrix renormalization group
(DMRG)\cite{Jiang}, matrix-product or tensor-network based
algorithms\cite{Murg,Yu,Furukawa,Wang}, high temperature
expansion\cite{Misguich}, resonating valence bond
approaches\cite{Capriotti2,Li,Beach,Hu}, exact solution\cite{Cai},
bond operator formalism\cite{Zhitomirsky,Ueda}, mean-field
theories\cite{Gelfand1,Mila,Isaev}, and field theoretical
methods\cite{Chandra2,Takano,Lante}. It has been established that in
the regime $0< J_{2}/J_{1} \lesssim 0.4$, the ground state of
$J_{1}-J_{2}$ model is an AFM phase with N$\acute{e}$el order. In
$J_{2}/J_{1} \gtrsim 0.6$, an AFM phase with collinear LRO is
stable, due to the dominance of the next-nearest-neighbor coupling
$J_{2}$. One of the most controversial regime is the intermediate
regime $0.4 \lesssim J_{2}/J_{1} \lesssim 0.6$ where the ground
state is non-magnetic and hence the SU(2) symmetry is not broken.
The nature of this intermediate non-magnetic ground state is still a
much debated issue. The possible candidates of this ground state, as
been proposed by various authors, include dimerized valence bond
solid (VBS) which breaks both the translation and the rotation
symmetries of the lattice\cite{Schulz,Gelfand1,Gelfand2,Singh1}, the
plaquette VBS which breaks only the translation
symmetry\cite{Zhitomirsky,Takano,Isaev}, the nematic spin liquid
which breaks only the rotational symmetry\cite{Lante}, and the
gapped\cite{Jiang,Wang,Li} or gapless\cite{Hu} spin liquid which
conserves all the symmetries of the lattice. The difficulty of this
issue lies in that there is no unbiased and accurate method to study
the ground state of $J_{1}-J_{2}$ model in the thermodynamical
limit. Most of the numerical studies heavily rely on the
extrapolation of the finite size results to the thermodynamical
limit. In cases where there is little guide from the analytical
knowledge, this practice may have uncertainties\cite{Matthieu,Yu} as
demonstrated by a recent study on the J-Q model\cite{Sandvik}.

Besides the nature of the nonmagnetic state, there are other
important issues under various physical contexts. Previous studies
show that AFM N$\acute{e}$el phase transits into the non-magnetic
state at $J_{2}/J_{1} \approx 0.4$ through a continuous quantum
phase transition. If the intermediate region actually possesses a
VBS order, this transition is an abnormal one, as a continuous
transition between two phases without the group-subgroup symmetries
violates the conventional "Landau rule". A "deconfined" quantum
critical point was proposed to exist between the N$\acute{e}el$ and
the VBS states\cite{Senthil}.

For the parameter regime $J_{2}/J_{1} \gtrsim 0.6$, this model also
invoked much interest since lots of real materials are related to
this parameter regime, such as the La-O-Cu-As iron based
superconductors\cite{Dai,Bruning} and $Li_2VOSiO_4$\cite{Melzi}.
Another interesting issue in this parameter regime is the possible
finite temperature symmetry breaking. For this model, although the
spin SU(2) symmetry cannot be broken spontaneously at finite
temperature due to the Mermin-Wagner theorem\cite{Mermin}, symmetry
breaking of the lattice $C_{4}$ symmetry could occur below a finite
$T< T_c$\cite{Chandra2,Weber,Capriotti3}. However, there is also a
different opinion on this issue\cite{Singh2}.

The effect of spin-anisotropy in the $J_{1}-J_{2}$ model is also an
interesting issue, given that the anisotropy is quite common in real
materials. Theoretical studies on this issue is
rare\cite{Viana,WangHY}.

In this paper, we focus on the phase boundary of the the
$J_{1}-J_{2}$ model and attempt to present accurate critical values
$J_{2}^{c1}$ and $J_{2}^{c2}$. We use the cluster mean-field theory
(CMFT), which is the cluster extension of the Weiss mean-field
theory\cite{Weiss1,Yamamoto}. We obtained the N$\acute{e}$el AFM
phase, the collinear AFM phase, and the nonmagnetic phase. Using the
reshaping method for plotting multiple-valued curves\cite{Tong}, we
studied the fine structure of the first order phase transition
between the nonmagnetic phase and the collinear AFM phases,
including the stable, meta-stable and unstable phases. These
informations are important when the system is under external
influence but are often neglected in previous studies. The critical
values $J_{2}^{c1}$ and $J_{2}^{c2}$ are found to converge very fast
with increasing cluster size, allowing us to obtain an accurate
estimation of them. We also analyze the finite temperature
properties, the mean-field results for which, though incorrect for
the isotropic model itself, are known to be relevant to the
corresponding properties of the anisotropic $J_{1}-J_{2}$ model.

The rest part of this paper is organized as follows: In Sec.
\uppercase\expandafter{\romannumeral2}, we introduce the CMFT and
the method we used to obtain the fine structure of the first-order
phase transition. In Sec. \uppercase\expandafter{\romannumeral3}, we
first present the zero temperature results in part A, including the
phase diagram and magnetic susceptibility. In part B, a phase
diagram at finite temperature is given and various susceptibilities
are presented and discussed.

\section{Method}

The simplest mean-field theory for spin systems is the Weiss's
single-site mean-field theory\cite{Weiss1}. In this theory, the
influence of surrounding spins to a central spin is approximated by
an effective static field, which is then determined
self-consistently. The Weiss mean-field theory thus neglects the
spatial fluctuations and often overestimates the stability of LRO.
Based on a similar idea, Bethe-Peierls-Weiss
(BPW)\cite{Bethe,Peierls,Weiss2} and Oguchi\cite{Oguchi} improved
the approximation by mapping the lattice model into clusters
subjected to self-consistently determined effective fields. The
interactions inside a cluster is treated exactly while interactions
between clusters are approximated by mean fields. Since the
short-range spatial fluctuations inside a cluster are taken into
account, the results are expected to improve as cluster size
increases.

In this work, we study the $J_{1}-J_{2}$ model on a square lattice
using the cluster extension of Weiss mean-field theory. Although
being simple, this theory produces surprisingly accurate boundaries
between various phases, as compared to results from more
sophisticated methods. We first divide the lattice into identical
clusters of $L$ sites. To separate the spin couplings inside a
cluster from those between clusters, the Hamiltonian of
$J_{1}-J_{2}$ model is rewritten as
\begin{eqnarray}
  \hat{H} &=& \sum_{c_{n}} \left[ J_{1} \sum_{\langle ij \rangle} \mathbf{S}_{i,c_{n}}  \cdot \mathbf{S}_{j,c_{n}}
  + J_{2} \sum_{\langle\langle ij \rangle\rangle}  \mathbf{S}_{i,c_{n}} \cdot  \mathbf{S}_{j,c_{n}} \right] \nonumber\\
   &+& \sum_{c_{n} \neq c_{m}} \left[ J_{1} \sum_{\langle ij \rangle}  \mathbf{S}_{i,c_{n}} \cdot  \mathbf{S}_{j,c_{m}}
   + J_{2} \sum_{\langle\langle ij \rangle\rangle}  \mathbf{S}_{i,c_{n}} \cdot  \mathbf{S}_{j,c_{m}} \right]. \nonumber\\
\end{eqnarray}
The operator $S_{i,c_{n}}$ donates the spin operator on the $i$-th
site in the cluster $c_{n}$. The first term in Eq.(2) represents the
Hamiltonian of decoupled clusters, while the second one represents
interactions between clusters. We make the standard mean-field
approximation for the interactions between two spins belonging to
different clusters $c_n \neq c_m$,
\begin{equation}
     \mathbf{S}_{i,c_{n}} \cdot  \mathbf{S}_{j,c_{m}} \approx S^{z}_{i,c_{n}} \langle S^{z}_{j,c_{m}} \rangle
      + \langle S^{z}_{i,c_{n}}\rangle S^{z}_{j,c_{m}} - \langle S^{z}_{i,c_{n}} \rangle \langle S^{z}_{j,c_{m}}
      \rangle.
\end{equation}
Here, $z$-axis is chosen as the quantization axis. This
approximation breaks both spin SU(2) symmetry and spatial
translation symmetry of the original Hamiltonian. Substituting it
into the second term of Eq.(2) and neglecting a constant, we obtain
the cluster-decoupled mean-field Hamiltonian,
\begin{eqnarray}
  \hat{H}_{mf} &=& \sum_{c_{n}} \hat{H}_{c_n} \nonumber \\
  \hat{H}_{c_n} &=& J_{1} \sum_{\langle ij \rangle} \mathbf{S}_{i,c_{n}} \cdot \mathbf{S}_{j,c_{n}}
  + J_{2} \sum_{\langle \langle ij \rangle \rangle} \mathbf{S}_{i,c_{n}} \cdot \mathbf{S}_{j,c_{n}} \nonumber \\
  && + \sum_{i=1}^{L} h_{i} S^{z}_{i,c_{n}}.
\end{eqnarray}
Here $h_{i}$ is the effective static field felt by the spin
$\mathbf{S}_{i,c_n}$. It is a linear combination of ${\langle
S^{z}_{j,c_m} \rangle}$, the magnetization of boundary site $j$ on
the neighboring cluster $c_m$.
\begin{figure}
\center
\includegraphics[width=250pt]{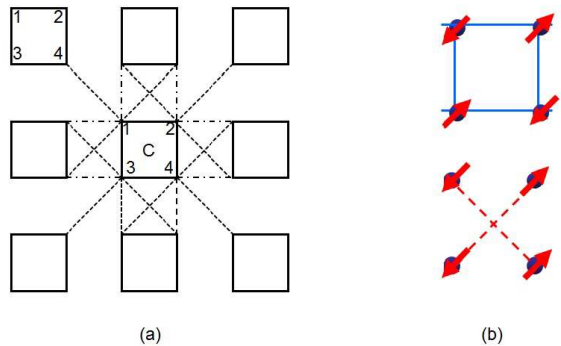}\\
\caption{(a) The square lattice is divided into $2 \times 2$
clusters (solid lines). The interactions between different clusters
are denoted by dot-dashed lines ($J_{1}$) and dashed lines
($J_{2}$). (b) Upper: picture of N$\acute{e}$el AFM order, dominated
by the nearest antiferromagnetic interaction $J_{1}$ (solid lines).
Lower: picture of collinear AFM order, dominated by the next nearest
antiferromagnetic interaction $J_{2}$ (dashed lines). }\label{Fig1}
\end{figure}

Fig.1 shows an example of $2 \times 2$ clusters and their couplings
between each other. We use the spatial translation symmetry of
clusters to ensure $\langle S^{z}_{i,c_{n}} \rangle = \langle
S^{z}_{i} \rangle = m_{i}$. For a cluster with $L$ sites, $m_{i}$
($i=1,2,...,L$) are our magnetic order parameters that can
characterize different magnetic orders. In this paper, we do not
consider the possibility of LRO in the intermediate non-magnetic
regime, as it is still an open issue how to incorporate the
non-magnetic order parameters into the CMFT. With this notation, the
effective field ${h_{i}}$ reads
\begin{equation}
   h_{i} = J_{1} \sum_{\delta} m_{\delta}  + J_{2} \sum_{\delta^{\prime}
   } m_{\delta^{\prime}}.
\end{equation}
Here $\delta$, $\delta^{\prime}$ $\in [1, L]$, denoting the nearest
neighbor site and the next-nearest neighbor site in the neighboring
clusters of site i, respectively. The CMFT equations are completed
by solving $m_{i}$ from a central cluster Hamiltonian $\hat{H}_{c}$
in Eq.(4). In the limit of single-site cluster $L=1$, the above
approximation recovers the Weiss mean-field theory. As the cluster
size increases, longer and longer range correlations contained in
the cluster are treated exactly. Therefore, the results are expected
to become exact as $L$ tends to infinity.

To solve the CMFT equations, we use open boundary conditions for the
cluster. The $L$ magnetization values $m_{i}$ ($i=1,2,...,L$) are
solved independently without symmetry constraints. Due to the lack
of translation symmetry within the cluster, $|m_i|$ has a weak
site-dependence, being smaller on the center of the cluster, and
larger on the edge and even larger at the corner. The qualitative
behavior of magnetization on different sites are exactly the same,
i.e. they will be zero or non-zero at the same time, indicting the
appearance or disappearance of the magnetic LRO.

We use iterative method to solve the mean-field equations. For a
given set of effective fields ${h_i}$, we use Lanczos method (for
$T=0$) or full ED method (for $T>0$) to calculate the magnetization
${m_i}$ which are feed back to Eq.(5). This process iterates until
all the $m_i$'s converge. For a given $J_{2}$ and $T$, the
calculation starts from a initial set of $m_i$'s, which we usually
get from the self-consistent solution of a slightly deviated
parameter $J_2$ (or $T$). Thus we can scan the parameter space from
small $J_2$ (or $T$) to larger values, or vice versa. It turns out
that the set of mean-field equations has more than one solutions,
stabilized respectively by scanning from left to right or from right
to left along the $J_2$ (or $T$) axis. For those multiple solutions
at a fixed ($J_2$, $T$), we compare their energies ($T=0$) or free
energies ($T>0$) to determine the physical solution of this system.
After the solutions of $m_i$ ($i=1,2,...,L$) are obtained, its LRO
can be identified easily from the magnetization pattern.

Near $J_{2}\approx 0.6$, naive scanning of $J_{2}$ produces a
discontinuous $m - J_{2}$ curve: $m$ jumps from $0$ to a finite
value or vice versa ($m$ is the magnetization of a center site of
the cluster). We suppose that this is the numerical instability due
to the multiple-valued relation of $m - J_{2}$. If such structure
does exist, ordinary calculation can only produce one branch of
solution and neglect the others, leading to a jump at some $J_{2}$
where the relative stabilities of two solutions invert. To overcome
this problem, we use the "stretching trick" proposed in the study of
first-order phase transitions in correlated electron
systems\cite{Tong}. If the mean-field solution $m = F(J_{2})$ is a
continuous curve in the $m - J_{2}$ plane but has a $S$- or
$Z$-shaped turn, the new equation $m = F(J_{2}-V |m|)$ will produce
a single-valued $m-J_2$ curve, given a proper selection of $V > 0$.
Pictorially this single-valued curve is obtained by "stretching" the
original curve. We can then solve this modified equation first and
recover the original solutions by plotting $m$ versus $J_{2}-V |m|$.

\section{Results and Discussions}

\subsection{Zero Temperature}
\begin{figure}
  \center
  \includegraphics[width=200pt, height=270pt, angle=270]{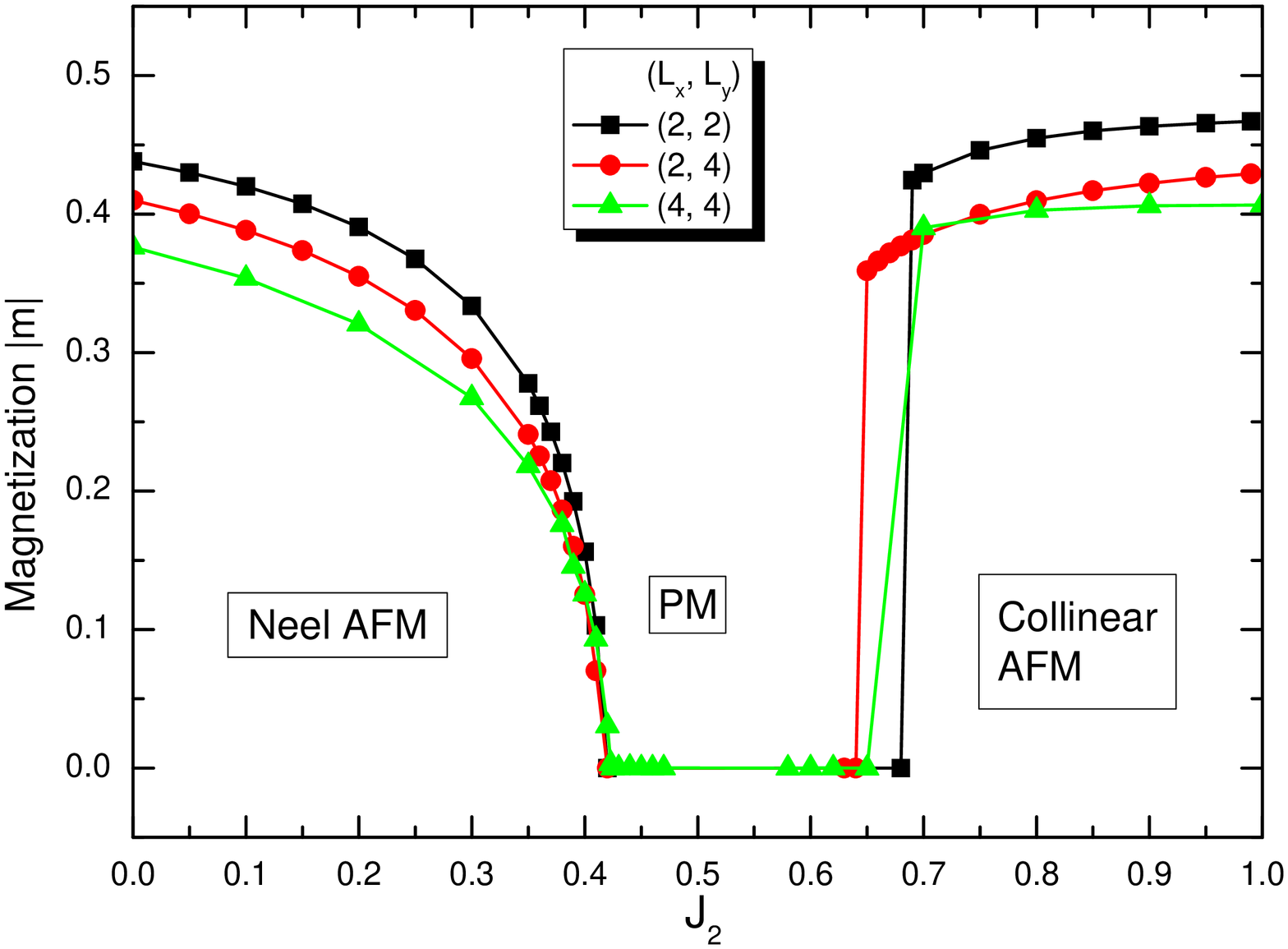}\\
  \caption{Magnetization $m$ versus $J_2$ obtained using various cluster geometries.
   The pattern of LRO's are marked in the figure. PM denotes paramagnetic.
   Here $m$ is the magnetization of a spin at the center of the cluster.}\label{Fig2}
\end{figure}
In this work, we use the rectangular clusters of size $L=L_{x}\times
L_{y}$. To avoid odd number of spins in a cluster, we use even $L_x$
and $L_y$. The total number of spins $L$ is confined as $L \leq 16$
due to the exponential increase of computational cost with $L$. We
choose $2 \times 2$ and $4 \times 4$ clusters for qualitative study,
and use $L_{y}=2$ and $L_{x}=2,4,6,8$ for quantitative size
dependence analysis.

In Fig.2, we show $|m|$ versus $J_{2}$ for three successively larger
clusters. $|m|$ is measured on the center site of the cluster. For
all the clusters we used, the N$\acute{e}$el order is stable for
small $J_{2}$ regime. As $J_{2}$ increases, $|m|$ decreases and
vanishes continuously at a critical value $J_{2}^{c1} \approx
0.41-0.42$, which indicates a second order transition to a
non-magnetic phase. As $J_{2}$ increases above $J_{2}^{c2} \approx
0.6 - 0.7$, $|m|$ jumps from zero to a finite value, with a
collinear magnetic pattern. In both N$\acute{e}$el and collinear
phases, $m$ decreases with increasing $L$, showing that more and
more quantum fluctuations are taken into account by using large
clusters, and hence the increasing quality of our results. The exact
value $m=0.307$\cite{Sandvik2} for $J_{2}=0$ is only asymptotically
approached in $L=\infty$ limit. It is interesting to observe that
the critical point $J_{2}^{c1}$ does not change much from $L=4$ to
$L=16$, showing that it converges very rapidly with $L$. Taking the
$L=16$ result as out estimation for the thermodynamical limit, we
obtain $J_{2}^{c1} \approx 0.42$. Compared to other methods such as
the ED\cite{Schulz, Richter}, series expansion\cite{Gelfand1,Singh1}
and DMRG\cite{Jiang}, CMFT is surprisingly accurate and simple in
producing the ground state phase boundaries.

\begin{figure}
\begin{center}
  \includegraphics[width=250pt, height=320pt, angle=270]{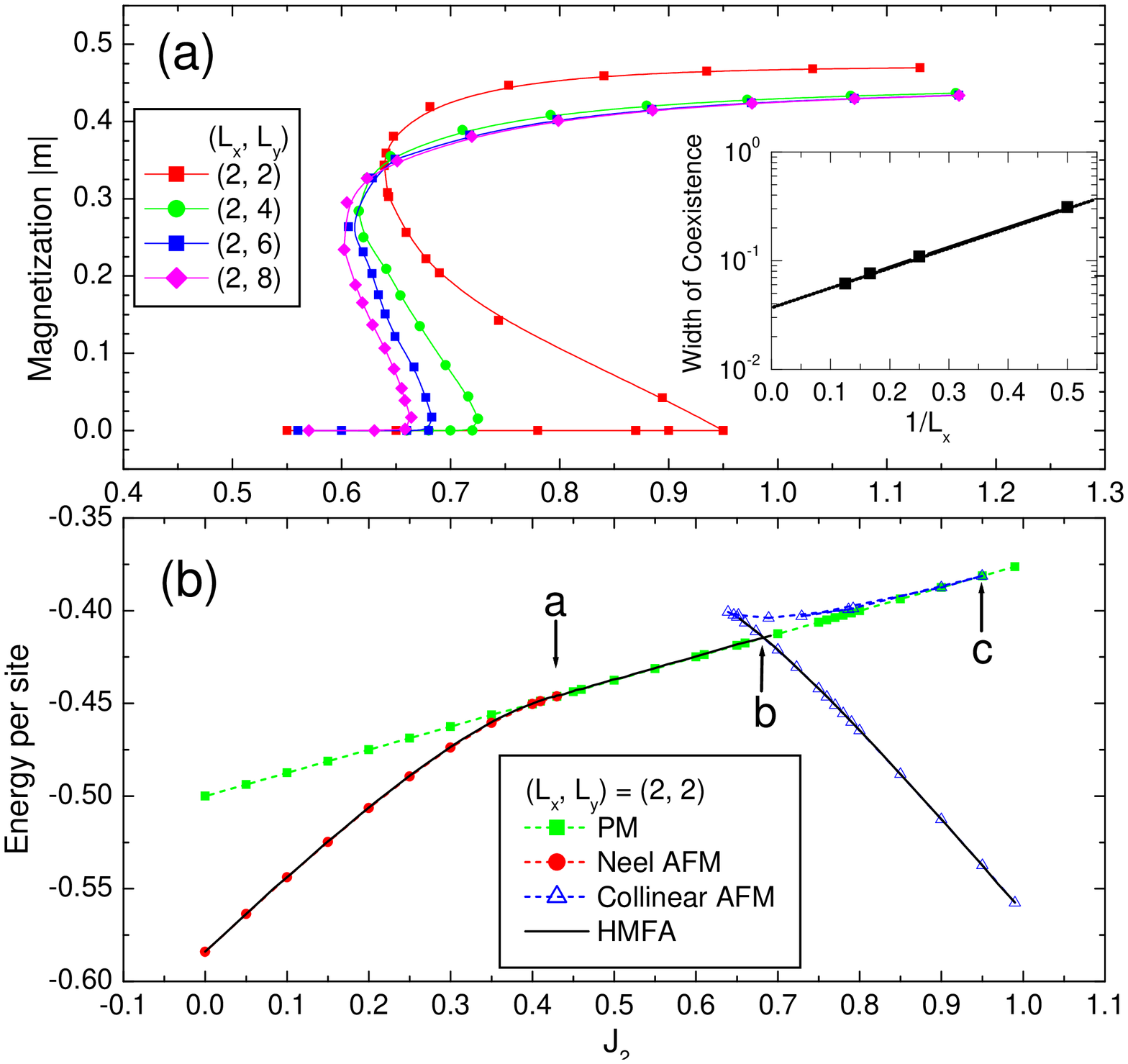}
  \vspace*{0.0cm}
\end{center}
  \caption{(a) Magnetization $|m|$ of a center spin versus $J_{2}$ near the first order phase transition for
  $L_{y}=2$ and $L_{x}=2$, $4$, $6$, and $8$, respectively. Inset: The width of coexistence
  region versus $1/L_{x}$. (b) Ground state energy per site versus
  $J_{2}$ in different phases, obtained using $L_{x}=L_{y}=2$
  cluster. The arrows mark the second-order N$\acute{e}$el-to-nonmagnetic
  transition (arrow a), the first order nonmagnetic-to-collinear transitions (arrow b),
  and the meta-stable second-order nonmagnetic-to-collinear transition (arrow c).
  Symbols are data and the dashed lines are for guiding the eyes.
  The solid line is the result of Hierarchical mean-field approach
  using $2 \times 2$ cluster in Ref.\cite{Isaev}.
}\label{Fig3}
\end{figure}

In Fig.3(a), we take a closer look at the fine structure of the
$|m|-J_{2}$ curve near $J_{2}^{c2}$, where the transition between
the non-magnetic phase and collinear AFM phase occurs. It is
obtained by the "stretching trick" mentioned above. In order to see
the systematic cluster size dependence, we fix $L_{x}=2$ and
increase $L_{y}$ from $2$ to $8$. We always obtain continuous curves
with $S$-shaped structures which contain the stable, meta-stable,
and the unstable phases and are generic features of the first order
phase transition. The width of the coexistence region $W$ decreases
as $L_{y}$ increases. As shown in the inset of Fig.3(a), $W$ is
found to scale with $1/L_{y}$ as $W \propto \alpha e^{\beta/ L_{y}}$
for the calculated cluster size. Fitting of the data gives $\alpha =
0.038$ and $\beta=0.42$. $\alpha =0.038 > 0$ means that the first
order phase transition still exists even if we use a cluster
$L_{x}=2, L_{y}=\infty$. This seems to be a strong support to the
first-order phase transition between non-magnetic phase and
collinear AFM phase in the thermodynamical limit. For a more
convincing conclusion, one should extrapolate $L_{x}$ and $L_{y}$ to
infinity simultaneously. However, due to the rapid increase of the
numerical cost, this is not done in our present study.

In Fig.3(b), the ground state energy per site versus $J_{2}$ is
plotted for the N$\acute{e}$el AFM, non-magnetic, and the collinear
AFM phases. We show the result obtained using $2 \times 2$ cluster
for demonstration purpose. As $J_{2}$ increases up to $0.42$ (marked
by arrow "a"), the energy of N$\acute{e}$el AFM continuously
approaches that of the non-magnetic phase from below, consistent
with the scenario of a second-order transition. The transition
between the non-magnetic phase and the collinear AFM phase occurs at
the energy crossing point marked by the arrow "b" in Fig.3(b), which
we denote as $J_{2}^{c2}$. In the coexistence region, a third
collinear AFM solution has the highest energy. It corresponds to the
unstable solution with negative $|m|-J_{2}$ slope in Fig.3(a). In
this first-order transition, a continuous transition does exist at
the meta-stable level, between collinear AFM and non-magnetic phases
(marked by arrow "c").

This scenario is common in first order phase transitions described
by mean-field equations, as disclosed by the dynamical mean-field
theory study for the correlated electron systems\cite{Tong}.
Extrapolating $L_{y}$ to infinity, we get $J_{2}^{c2} \approx 0.59$,
which should be very close to the exact value in the thermodynamical
limit. This value agrees quite well with the more sophisticated
calculations such as DMRG\cite{Jiang} (see Table.1 below). It is
noted that our energy curve agree quantitatively with the result
from the hierarchical mean-field approach (HMFA) on $2 \times 2$
cluster\cite{Isaev} (solid lines in Fig.3(b)). Although HMFA is
based on the sophisticated Schwinger boson representation and
mean-field approximation, the quantitative agreement makes us
believe that the HMFA is equivalent to the cluster mean-field method
that we used here, at least for the case of $2 \times 2$ cluster.
The critical values of $J_{2}$ have been obtained in many works,
using different methods with varied sophistications. In Table.1, we
summarize some of the previous results and compare them with ours.
Note that a similar CMFT study on the $J_1-J_2$ model was carried
out in Ref.\cite{Gelfand1}, but the cluster size effect was not
analyzed systematically.

\begin{figure}
\begin{center}
  \includegraphics[width=200pt, height=270pt, angle=270]{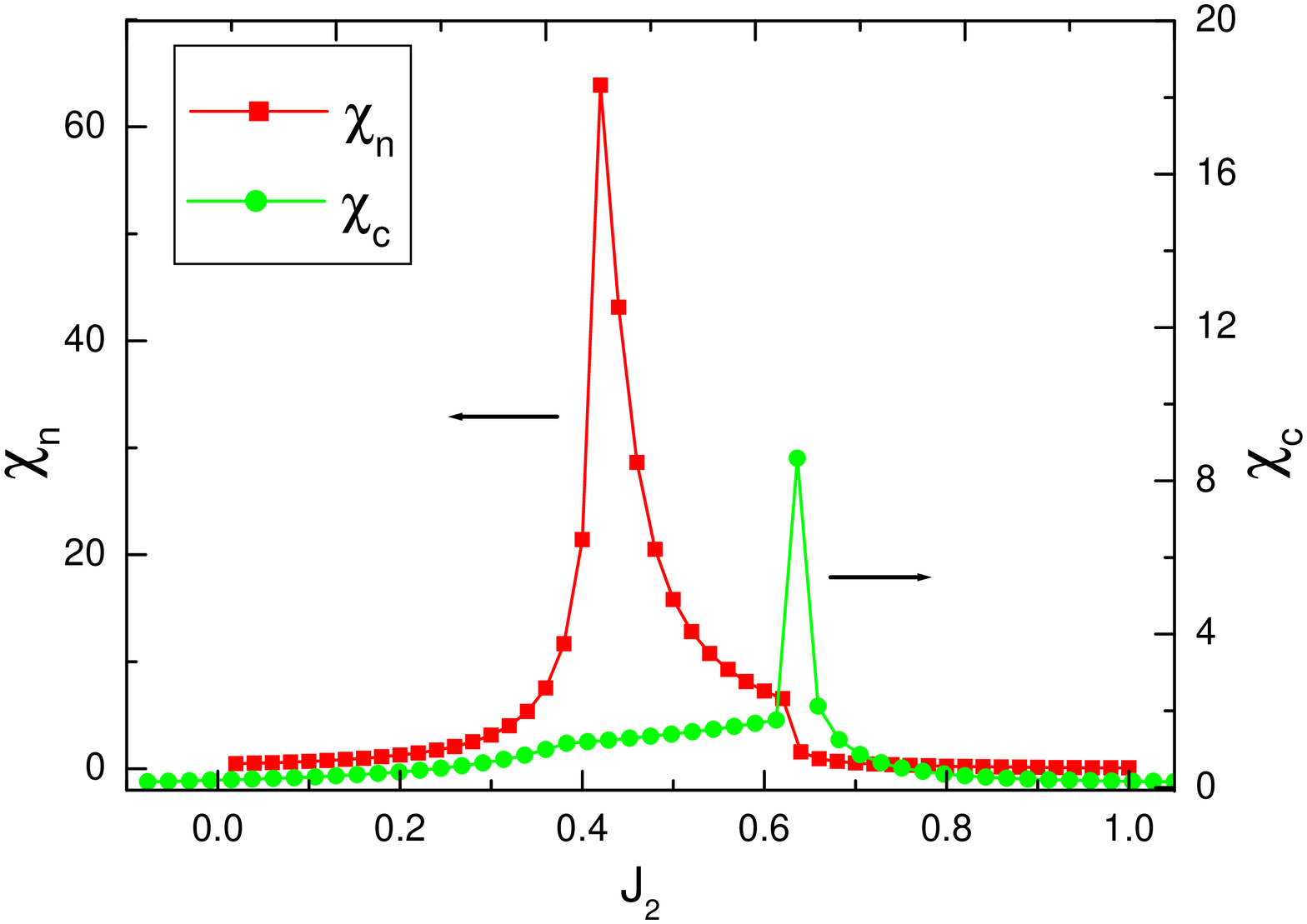}\\
\vspace{-1.0cm}
\end{center}
  \caption{ Zero temperature N$\acute{e}$el susceptibility $\chi_{n}$ (squares with guiding line)
  and collinear susceptibility $\chi_{c}$ (dots with guiding line) as functions of $J_{2}$.
The data are obtained by numerical derivation with the applied field
$h=0.01$. }\label{Fig4}
\end{figure}
%

%
\begin{table}
\caption{Comparison of $J_{2}^{c1}$ and $J_{2}^{c2}$ from various
works. The methods are abbreviated as ED(exact diagonalization),
SE(series expansion), DMRG(density-matrix renormalization group),
HMFT(hierarchical mean-field theory), VMC(variational Monte Carlo),
and CMFT(cluster mean-field theory).}
\center
\label{tab:1}       
\begin{tabular}{llllllll}
\hline\noalign{\smallskip} Ref. & \hspace{0.1cm} \cite{Richter}
\hspace{0.0cm} &  \hspace{0.0cm} \cite{Gelfand1} & \hspace{0.0cm}
\cite{Jiang}
&\hspace{0.0cm} \cite{Darradi} &\hspace{0.0cm} \cite{Isaev} &\hspace{0.0cm} \cite{Hu} & \hspace{0.0cm} this work  \\
Met. & \hspace{0.0cm} ED \hspace{0.0cm} & \hspace{0.0cm} SE &
\hspace{0.0cm} DMRG &\hspace{0.0cm} CC &\hspace{0.0cm} HMFT &\hspace{0.0cm} VMC & \hspace{0.0cm} CMFT  \\
 \noalign{\smallskip}\hline\noalign{\smallskip}
$J_{2}^{c1}$ & \hspace{0.0cm} 0.35  & \hspace{0.0cm} 0.41  & \hspace{0.0cm}  0.41  & \hspace{0.0cm}  0.44  & \hspace{0.0cm} 0.42  & \hspace{0.0cm} 0.45 & \hspace{0.0cm} 0.42\\
$J_{2}^{c2}$ & \hspace{0.0cm} 0.66  & \hspace{0.0cm} 0.64  & \hspace{0.0cm}  0.62  & \hspace{0.0cm}  0.59  & \hspace{0.0cm} 0.66  & \hspace{0.0cm} 0.6  & \hspace{0.0cm} 0.59\\
\noalign{\smallskip}\hline
\end{tabular}
\end{table}

A central issue in the study of $J_{1}-J_{2}$ model is the
properties of the intermediate non-magnetic phase. The key question
is whether it is a spin liquid or a VBS that breaks the lattice
translation and/or rotation symmetry. Since in CMFT, the translation
symmetry of the original lattice is broken by hand, we cannot answer
this question directly. In the non-magnetic phase, the effective
fields of CMFT become zero and $H_{mf}$ describes uncorrelated
clusters. Then CMFT is equivalent to the bare ED on a cluster with
open boundary condition, in contrast to periodic boundary condition
commonly used in previous ED studies. The open boundary condition
will induce nonzero VBS order parameter in small clusters. For an
example, the operator of plaquette order parameter reads\cite{Fouet}
\begin{eqnarray}
  Q_{\alpha\beta\gamma\delta} &=& 2[(\mathbf{S}_{\alpha}\cdot \mathbf{S}_{\beta})(\mathbf{S}_{\gamma}\cdot \mathbf{S}_{\delta})
   +(\mathbf{S}_{\alpha}\cdot \mathbf{S}_{\delta})(\mathbf{S}_{\beta}\cdot \mathbf{S}_{\gamma}) \nonumber \\
   &&-(\mathbf{S}_{\alpha}\cdot \mathbf{S}_{\gamma})(\mathbf{S}_{\beta}\cdot \mathbf{S}_{\delta})]
   + \frac{1}{2} ( \mathbf{S}_{\alpha}\cdot \mathbf{S}_{\beta}+ \mathbf{S}_{\gamma}\cdot \mathbf{S}_{\delta}\nonumber \\
   &&+ \mathbf{S}_{\alpha}\cdot \mathbf{S}_{\delta}+ \mathbf{S}_{\beta}\cdot \mathbf{S}_{\gamma}
   + \mathbf{S}_{\alpha}\cdot \mathbf{S}_{\gamma}+ \mathbf{S}_{\beta}\cdot
   \mathbf{S}_{\delta} + \frac{1}{4} ) .  \nonumber \\
\end{eqnarray}
Here $\alpha, \beta, \gamma, \delta$ denote the four sites of a
plaquette clockwise. At $J_{2}=0.5$, the plaquette order parameter
is evaluated on a $2 \times 2$ cluster as
$Q_{\alpha\beta\gamma\delta}\thickapprox 0.988$, very close to its
saturate value $1.0$. Evaluating $Q_{\alpha\beta\gamma\delta}$ on a
larger cluster also gives nonzero result. However, these are the
boundary effect of the cluster and does not support a true VBS
state. It is an interesting open question how to incorporate the
order parameter of various VBS state into the mean-field
approximation. If such a mean-field theory does exist, considering
that it tends to exaggerated the LRO, a negative result about the
existence of VBS may rule out the possibility of VBS in the
intermediate parameter regime.

We also investigate the N$\acute{e}$el as well as collinear magnetic
susceptibility at zero temperature. These susceptibilities are
defined as
\begin{eqnarray}
  \chi_{\alpha} &=& \displaystyle\lim_{h\rightarrow 0^{+}} \frac{Tr \left[ e^{-\beta(\hat{H} - hM_{\alpha})}
  M_{\alpha} \right] }{Tr \left[ e^{-\beta(\hat{H} - hM_{\alpha})}\right] }.
\end{eqnarray}
Here, the N$\acute{e}$el susceptibility $\chi_{n}$ and collinear
susceptibility $\chi_{c}$ are defined using staggered magnetization
$M_{n}$ and $M_{c}$, respectively. For the $2 \times 2$ cluster
shown in Fig.1, $M_{n}=S_{1}^{z}-S_{2}^{z}+S_{3}^{z}-S_{4}^{z}$ and
$M_{c}=S_{1}^{z}+S_{2}^{z}-S_{3}^{z}-S_{4}^{z}$. We apply a small
staggered field $h$ and evaluate $\chi_{n}$ and $\chi_{m}$ using
numerical derivation. The results obtained are shown in Fig.4.

\begin{figure}
  \center
  \includegraphics[width=180pt, height=270pt, angle=270]{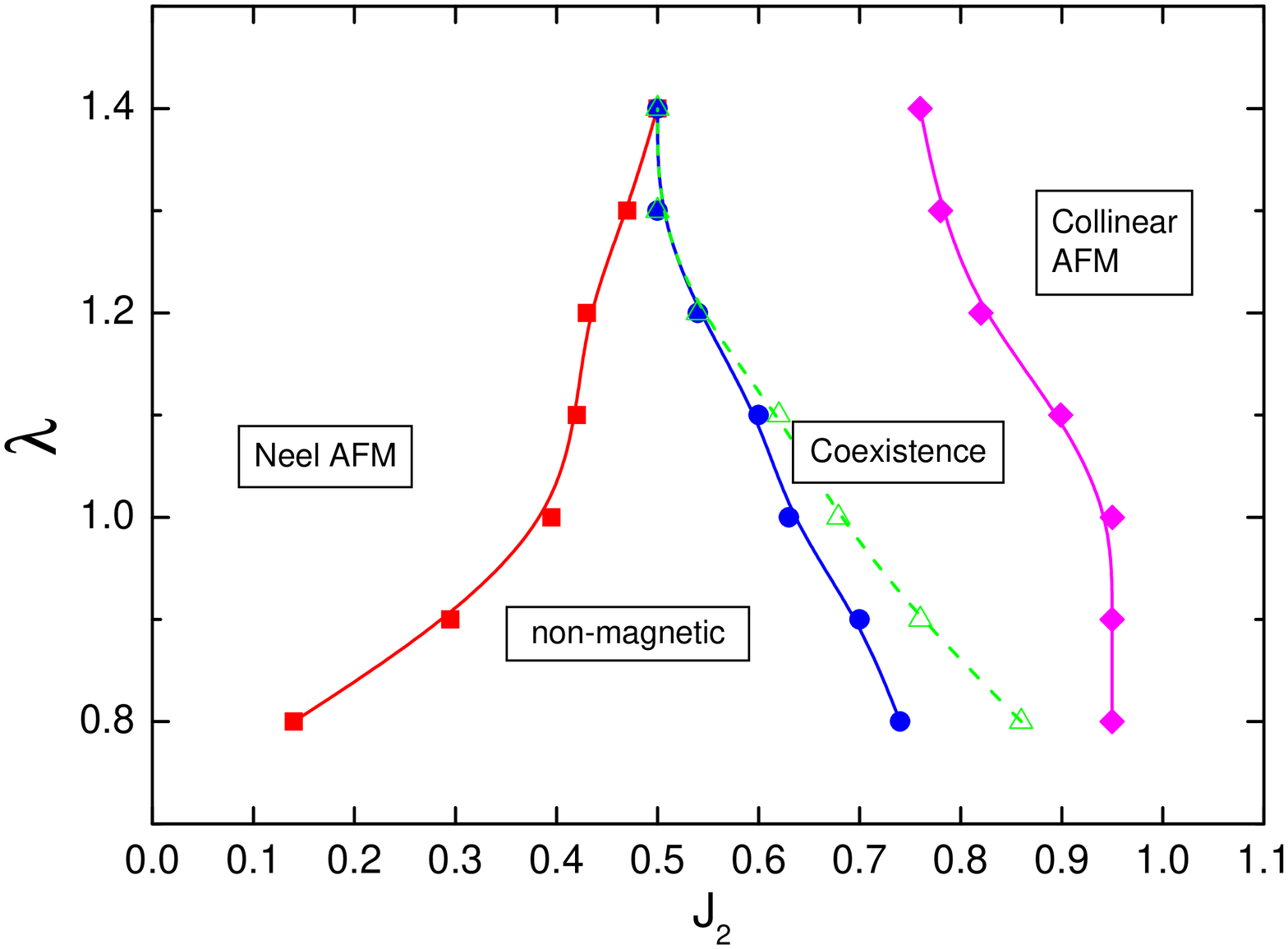}\\
  \caption{Phase diagram on the $\lambda-J_{2}$ plane obtained using $2 \times 2$ cluster.
  Squares with solid line is the second-order transition between
  N$\acute{e}$el AFM and non-magnetic phases. Dots and diamonds with solid
  lines represent the phase coexistence boundary of non-magnetic
  phase and the collinear phase. The triangle with dashed line is
  the actual first-order phase transition line.
  The lines are for guiding eyes.}\label{Fig5}
\end{figure}

The continuously diverging behavior of $\chi_{n}$ at $J_{2}\approx
0.42$ confirms the continuous transition from N$\acute{e}$el AFM
phase to non-magnetic phase. In contrast, near the collinear
transition $J_{2}^{c2}$, an abrupt jump of $\chi_{c}$ is observed,
being consistent with a first-order phase transition. Note that both
$\chi_n$ and $\chi_c$ are much larger in the non-magnetic regime
than in their corresponding long-ranged ordered regime. This shows
that the intermediate non-magnetic ground state is rich of short
range spin fluctuations at various momentums, and different types of
spin correlation compete strongly with each other. This leads to the
notorious difficulty in the study of the non-magnetic state.

The mean-field approximation used in our study introduces a symmetry
breaking term $H^{\prime} = \sum_{i=1}^{L} h_{i} S^{z}_{i,c_{n}}$,
which breaks the SU(2) symmetry of the original Hamiltonian. For
CMFT calculation using a finite cluster, this term effectively
suppresses the quantum fluctuation and tends to exaggerate the
stability of LRO in the ground state. As a result, the obtained
$|m|$ is larger than the exact value (as checked at $J_{2}=0$ case).
The region of the magnetic LRO is enlarged and non-magnetic region
suppressed. Here, to phenomenologically study the effects of
enhancing or reducing quantum fluctuations, we introduce artificial
fluctuations by multiplying a tunable factor $\lambda$ to the
mean-field term $H^{\prime}$. The total Hamiltonian becomes
$H_{eff}=H_{c_{n}}+\lambda H^{\prime}$. $\lambda < 1$ enhances the
fluctuation of $H_{eff}$, and it mimics the effects of larger
cluster or smaller $S$.  $\lambda > 1$ reduces the fluctuation of
$H_{eff}$ and it mimics the effects of anisotropy or larger spin.
Fig.5 shows a phase diagram in $\lambda-J_{2}$ plane. For larger
$\lambda$, the LRO region is enlarged and the non-magnetic region
shrinks. At $\lambda=1.4$, non-magnetic region diminishes, leading
to a direct first-order transition between N$\acute{e}$el phase and
collinear phase. At this point the phase diagram resembles that of
the $J_{1}-J_{2}$ Ising model where quantum fluctuation disappears.
For smaller $\lambda$, the non-magnetic region enlarges and for
sufficiently small $\lambda$, the LRO regime will disappear. This
phase diagram resembles the phase diagram of anisotropic Heisenberg
model\cite{Viana}. Note that this artificial fluctuation does not
influence the width of coexistence region, showing that the
first-order phase transition at $J_{2}^{c2}$ is robust against
quantum fluctuations.

\subsection{Finite Temperature}
\begin{figure}
\begin{center}
  \includegraphics[width=250pt, height=350pt]{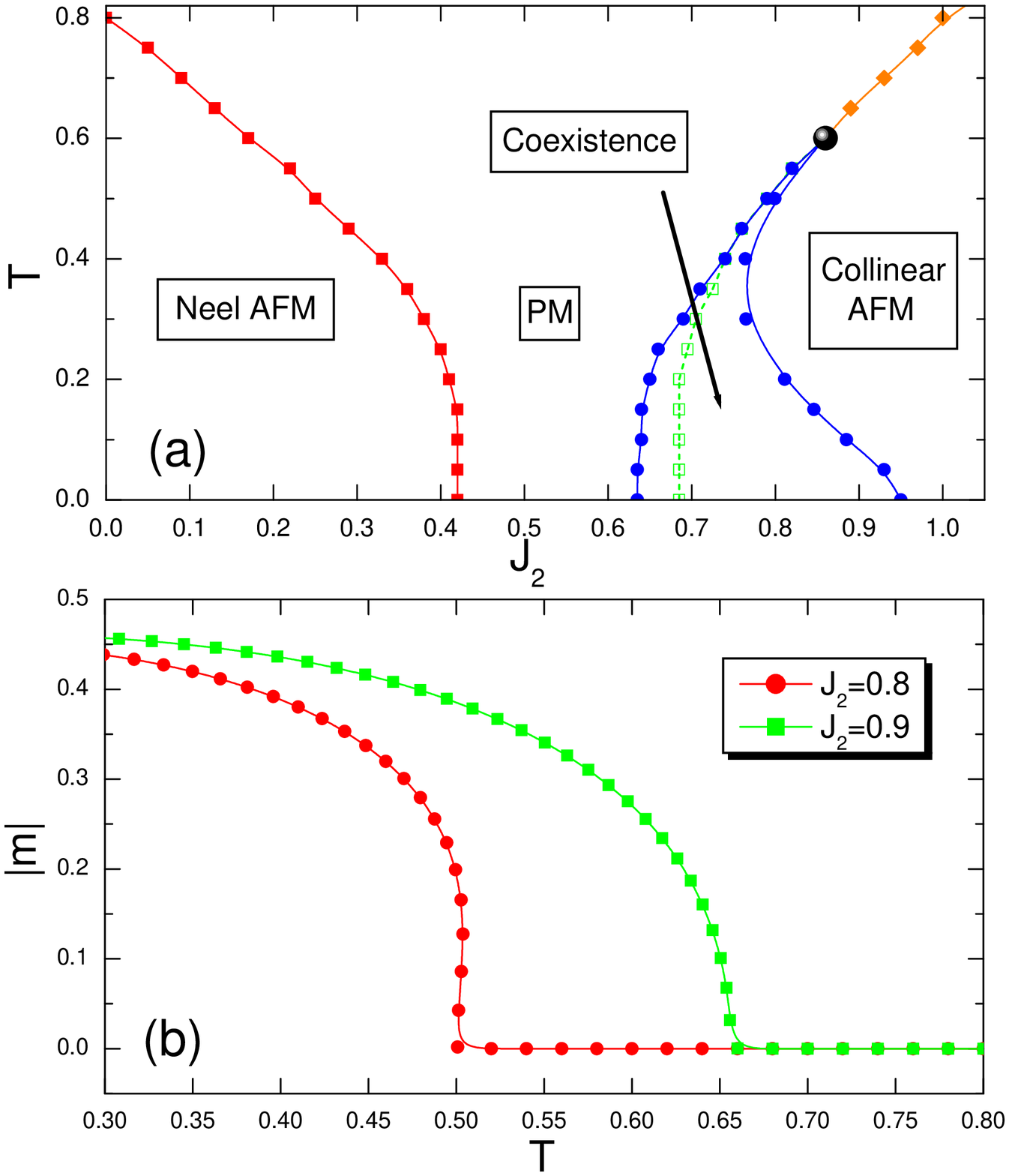}\\
\vspace{-3cm}
\end{center}
  \caption{(a) Phase diagram of $J_{1}$-$J_{2}$ model in the $T-J_{2}$ plane, obtained
  using $2 \times 2$ cluster mean-field theory. Squares with eye
  guiding line is the second-order N$\acute{e}$el-to-paramagnetic phase transition.
  Solid dots represent coexistence boundaries of paramagnetic phase and
  collinear AFM phase. The empty squares with dashed line is the actual transition line
  of equal free energy. The solid dot at ($J_{2c}=0.86$, $T_{c}=0.6$) is the critical point above which the
first-order transition changes into a second-order line (diamonds
with solid line). (b) magnetization $|m|(T)$ curves at $J_{2}=0.8 >
J_{2c}$ and $J_{2}=0.9 < J_{2c}$.}\label{Fig6}
\end{figure}

For finite temperatures, $J_{1}-J_{2}$ model does not have finite
magnetization, due to the Mermin-Wanger theorem. The mean-field
approximation used in CMFT suppresses the quantum fluctuations and
leads to a finite magnetization at $T>0$. $m$ approaches zero only
in the large $L$ limit. As a result, CMFT is not suitable for the
study of finite temperature properties of $J_1-J_2$ model in two
dimensions. Due to the effective suppression of quantum fluctuations
in CMFT, however, a finite cluster CMFT calculation for the
$J_1-J_2$ model can be used to qualitatively produce the phase
diagram of the spin-anisotropic $J_{1}-J_{2}$ model, such as the
$J_{1}^{xxz}-J_2$ model\cite{Viana}. In the following, we present
the finite temperature properties of the CMFT (using L=4), with the
possible relevance to the anisotropic $J_1-J_2$ model in mind.

Using ED method to solve the effective cluster Hamiltonian, we
obtain the $T-J_{2}$ phase diagram using $2\times 2 $ cluster as
shown in Fig.6(a). We scan along $J_{2}$ or $T$ axis to obtain the
full structure of the phase diagram. For $J_{2} < J_{2}^{c1} \approx
0.42$, there is a continuous transition line $T_{n}(J_{2})$
separating the low temperature N$\acute{e}$el state from the high
temperature paramagnetic phase. For $J_{1}-J_{2}$ model, the finite
$T_n$ is an artefact of the mean-field theory. As stated above,
however, it qualitative describes the trends of $T_n$ for the
anisotropic $J_{1}-J_{2}$ model. It is expected that $T_{n}$ tends
to zero in the limit of infinite cluster size. Indeed, using $2
\times 4$ cluster we obtain lower $T_{n}$. As $J_{2}$ increases,
$T_{n}$ decreases and vanishes at $J_{2} \approx 0.42$ continuously.

In the regime $J_{2}> 0.62$, at low temperatures, there is a finite
coexisting regime of the paramagnetic phase and the collinear AFM
phase. As temperature increases, this coexisting regime shrinks to a
point at $J_{2c} = 0.86$ and $T_{c}=0.6$. It is the critical point
separating the first-order phase transition and the second-order
transition. For $T>T_{c}$, the collinear-to-paramagnetic phase
transition becomes continuous. The whole phase diagram resembles the
that of the anisotropic $J_1-J_2$ model obtained using the effective
field theory\cite{Viana}. In Fig.6(b), two $|m|-T$ curves are shown
for $J_{2} = 0.8 < J_{2c}$ and  $J_{2} = 0.9 > J_{2c}$,
respectively. For $J_{2} = 0.8$, the $|m|-T$ curve has a slight
multiple-value region, corresponding to a weak first-order phase
transition. While for $J_{2} = 0.9$, it is a second-order phase
transition. In creasing the cluster size, we observe that the
transition temperature decreases.

For the $J_{1}-J_{2}$ model, a finite temperature phase transition
in regime $J_2 > J_2^{c2}$ may exist to break the $C_4$ rotation
symmetry of the lattice, according to Chandra {\it et
al.}\cite{Chandra2,Weber,Capriotti3}. However, what we obtained in
Fig.6(b) is nothing to do with this transition. It would be
interesting to develop our CMFT for further study of this novel
Ising transition. We leave this issue for the future.

\begin{figure}
  \center
  \includegraphics[width=200pt, height=270pt, angle=270]{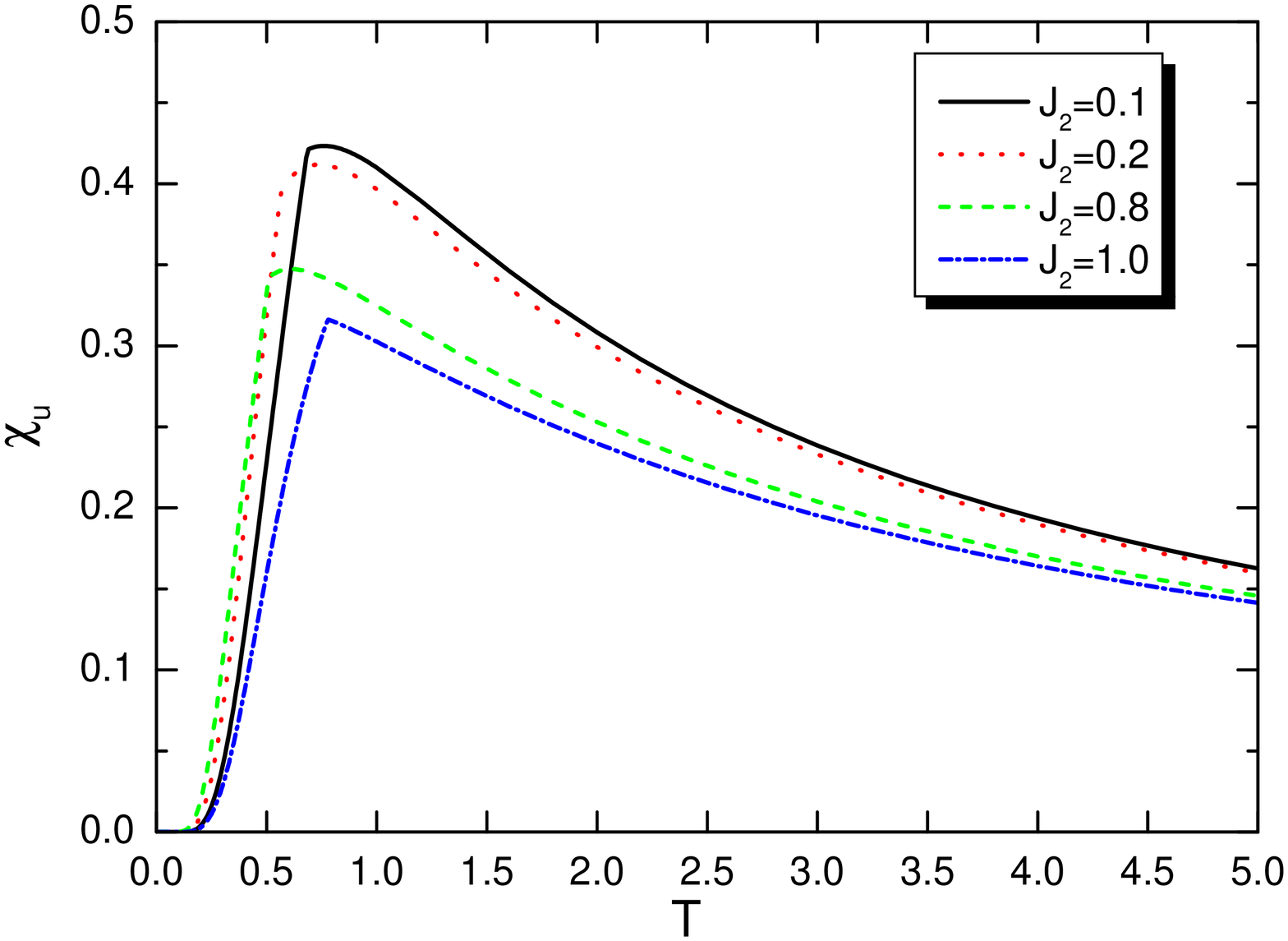}\\
  \caption{Uniform magnetic susceptibility $\chi_{u}$ versus $T$ for various $J_{2}$ values shown in the figure.
Inset: the maximum height of $\chi_{u}(T)$ as a function of $J_{2}$.
It is obtained using $2 \times 2$ CMFT.}\label{Fig7}
\end{figure}
In the end, we calculate magnetic susceptibilities as functions of
temperature. The uniform susceptibility $\chi_{u}$ (shown in Fig.7)
obeys Curie-Weiss law at high temperatures. For any value of $J_{2}$
that we studied, $\chi_{u}$ reaches zero exponentially in the $T=0$
limit, forming a peak at some finite temperature. The disappearance
of $\chi_{u}$ at $T=0$ shows that there is a finite gap in the
magnetic excitation. This may be an artefact due to the small
cluster that we used as well as due to the mean-field approximation.
At the transition temperature, a cusp in $\chi_{u}(T)$ is observed,
reflecting the singularity at the phase transition. In Fig. 8,
N$\acute{e}$el staggered susceptibility $\chi_{n}$ and collinear
staggered susceptibility $\chi_{c}$ are shown for $J_{2}=0.2$ and
$0.8$. The divergences in $\chi_{n}(T)$ for $J_{2}=0.2$ and in
$\chi_{c}(T)$ for $J_{2}=0.8$ are consistent with the finite
temperature transition, while $\chi_{n}(T)$ for $J_{2}=0.8$ and
$\chi_{c}(T)$ for $J_{2}=0.2$ only show a cusp or kink at the
transition temperatures.
\begin{figure}
\begin{center}
  \includegraphics[width=200pt, height=270pt, angle=270]{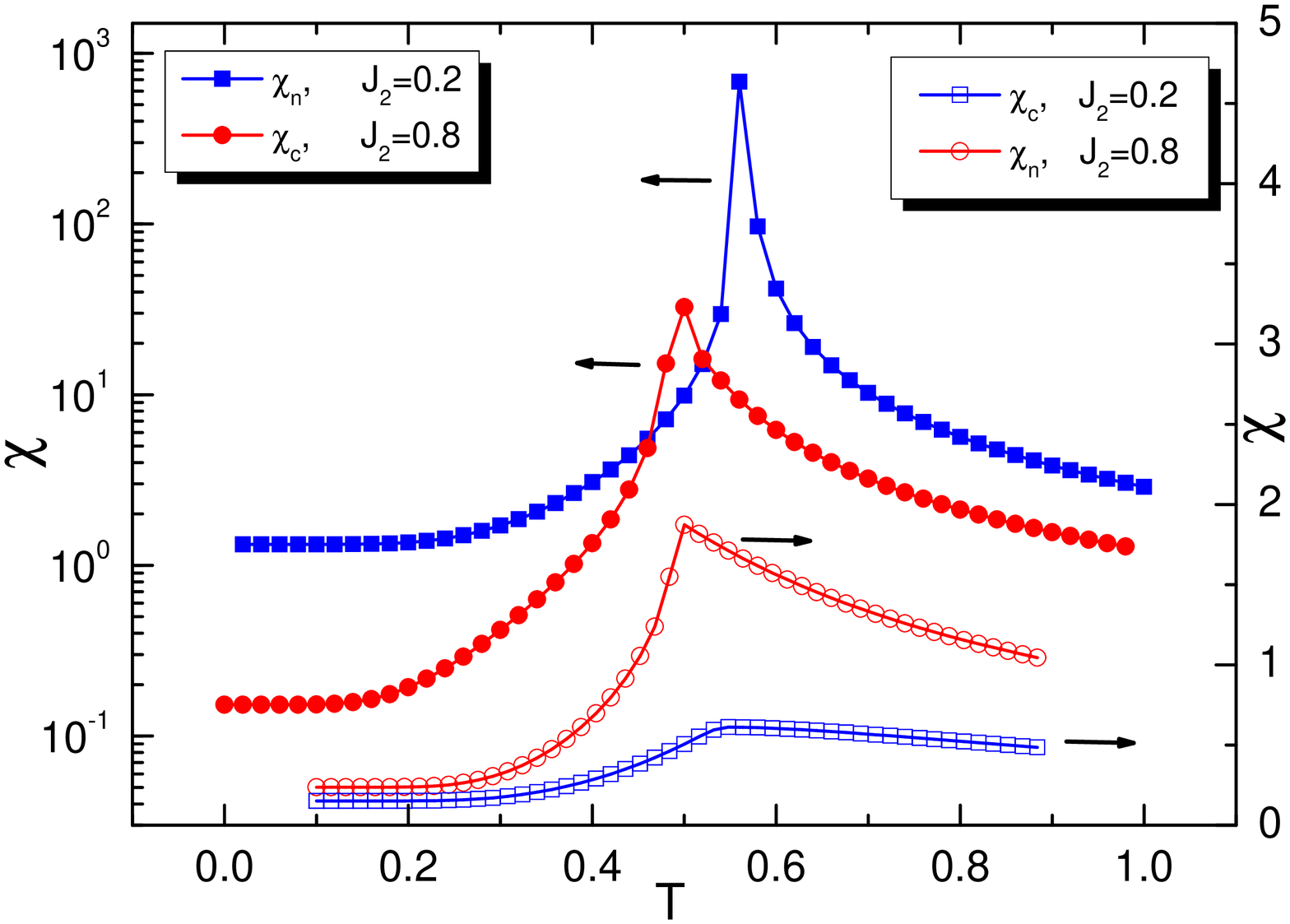}\\
\vspace{-1.0cm}
\end{center}
  \caption{The solid squares and dots with eye-guiding lines are $\chi_{n}$ at $J_{2}=0.2$ and
 $\chi_{c}$ at $J_{2} =0.8$, respectively. They show divergence at the transition temperature.
 The empty squares and dots with eye-guiding lines are  $\chi_{c}$ at $J_{2}=0.2$ and
 $\chi_{n}$ at $J_{2} =0.8$, respectively.}\label{Fig8}
\end{figure}

\section{Summary}
In summary, we use the cluster mean-field theory to study the
$J_{1}$-$J_{2}$ Heisenberg model on a square lattice. For small,
intermediate, and large $J_{2}/J_{1}$ regime, we obtain the
N$\acute{e}$el AFM phase, the non-magnetic phase, and the collinear
AFM phase, respectively. The N$\acute{e}$el-to-non-magnetic
transition is found to be of second order, and the
non-magnetic-to-collinear transition is of first order. The
respective critical values $J_{2}^{c1}$ and $J_{2}^{c2}$ are found
to converge rapidly with increasing $L$. From the largest $4\times
4$ cluster we obtain obtain $J_{2}^{c1} \approx 0.42$, which is very
close to the results of $2 \times 2$ cluster $0.41$. Extrapolating
the cluster size to infinity, we obtain $J_{2}^{c2} \approx 0.59$.
Both $J_{2}^{c1}$ and $J_{2}^{c2}$ agree with the previous results
very well. We also investigate the finite temperature phase diagram,
which due to the mean-field approximations, resembles that of the
anisotropic $J_{1}-J_{2}$ model. The first order transition in
$J_{2} > J_{2}^{c2}$ regime changes into a second order transition
at $T > T_c$. Various susceptibilities are discussed to help us
understand the system's behavior near critical point. Our results
show that the cluster mean-field theory is not only a very useful
tool for studying classical phase transitions\cite{Yamamoto}, but
can also give surprisingly accurate ground state phase boundaries
for the frustrated quantum magnet.

\section{Acknowledgement}
This work is supported by National Program on Key Basic Research
Project (973 Program) under Grant No. 2009CB29100, 2012CB821402, and
2012CB921704, and by the NSFC under Grant No. 91221302 and 11074302.

\end{document}